\documentstyle[12pt]{article}
\begin{document}
\voffset -0.7 true cm
\hoffset 1.1 true cm
\topmargin 0.0in
\evensidemargin 0.0in
\oddsidemargin 0.0in
\textheight 8.6in
\textwidth 7.1in
\parskip 10 pt

\newcommand{\be}{\begin{equation}}
\newcommand{\ee}{\end{equation}}
\newcommand{\bea}{\begin{eqnarray}}
\newcommand{\eea}{\end{eqnarray}}
\newcommand{\beas}{\begin{eqnarray*}}
\newcommand{\eeas}{\end{eqnarray*}}

\def\kl{{\frac{2 \pi l}{\beta}}}
\def\km{{\frac{2 \pi m}{\beta}}}
\def\kn{{\frac{2 \pi n}{\beta}}}
\def\kr{{\frac{2 \pi r}{\beta}}}
\def\ks{{\frac{2 \pi s}{\beta}}}
\def\b{{\beta}}
\font\cmsss=cmss8
\def\C{{\hbox{\cmsss C}}}
\font\cmss=cmss10
\def\bigC{{\hbox{\cmss C}}}
\def\scriptlap{{\kern1pt\vbox{\hrule height 0.8pt\hbox{\vrule width 0.8pt
  \hskip2pt\vbox{\vskip 4pt}\hskip 2pt\vrule width 0.4pt}\hrule height 0.4pt}
  \kern1pt}}
\def\ba{{\bar{a}}}
\def\bb{{\bar{b}}}
\def\bc{{\bar{c}}}
\begin{titlepage}
\begin{flushright}
{\small hep-th/0302012}
\end{flushright}

\begin{center}

\vspace{2mm}

{\Large \bf Cut-off QFT and black hole entropy}

\vspace{3mm}

Gilad Lifschytz 

\vspace{1mm}
{\small \sl Department of Mathematics and Physics} \\
{\small \sl and CCMSC}\\
{\small \sl University of Haifa at Oranim, Tivon 36006, ISRAEL} \\
{\small \tt giladl@research.haifa.ac.il}

\end{center}

\vskip 0.3 cm

\noindent
We clarify the relationship between black hole entropy and the number of degrees of 
freedom in the dual QFT with a cut-off. We show that simple gravity arguments predict
the correct cut-off procedure.

\end{titlepage}
\section{Introduction}
As is well known the entropy of a black hole is \cite{bh}
\begin{equation}
S_{BH}=\frac{A}{4G}
\end{equation}
where $A$ is the area of the horizon and $G$ is Newton constant.
This has led to the idea of holography \cite{hol}, which claims that the number of degrees of freedom
in a gravitational system is proportional to its area and not its volume, and that black hole 
configurations maximize the entropy for a given energy and charge.

In the context of $AdS/CFT$ \cite{mald} this takes a more precise form.
There is a connection between a CFT with a cut-off (to be defined 
later) and portions of AdS \cite{suswit} . The  full CFT lives at the boundary of AdS and is dual to string theory 
in the whole bulk of AdS. If one considers a CFT with some cut-off then one can think of
the cut-off CFT as `living'' on some 
time-like  hyper-surfaces in AdS, and this cut-off CFT is dual to string theory
in the inner bulk.  
Holography then implies that the number of degrees of freedom of the cut-off 
CFT (to be defined), related to some constant radial hyper-surface, is 
exactly the area of the hyper-surface in the AdS space divided by $4G$. 
In \cite{suswit}  
some definitions of the
number of degrees of freedom of a cut-off CFT were used and it was shown that it is 
proportional to the area in Planck units but they  have not specified the
regularization scheme that one should use.

Three quantities; The black hole entropy, the area in Planck units and the number of degrees of freedom in a cutoff CFT need to match to each other. Given the $AdS/CFT$
one should be able to understand why these quantities are
equal. Recently the
relationship between black hole entropy and the area in Planck units was explained in this context \cite{ikll2}.
In this note we would like to give a better understanding of the equality between the black hole entropy, which is the dual QFT thermal entropy and the number of degrees of freedom of the QFT with some cutoff.

\section{Thermality and cut-off}

Let us start with a CFT and label the number of states with energy $E$ by $\Omega(E)$.
The total number of states in a local quantum field theory without a cut-off
\begin{equation}
\int_{-\infty}^{\infty}\Omega(E) dE
\label{states}
\end{equation}
is of course infinite. At any given energy one could have defined (ad hoc)
the number of degrees of freedom to be
\begin{equation}
\ln \Omega(E)
\end{equation}
If one looks at a quantum field theory with some cut-off scheme then one 
can define a regulated version of (\ref{states})
\begin{equation}
N(\Lambda)=\int f(\Lambda, E) \Omega(E)dE
\end{equation}
where $f(\Lambda,E)$ is a regulating function defining the regularization 
scheme.
How should one define then 
the regulated number of degrees of freedom ?. Since $\Omega(E)$ is a 
rapidly increasing function (at least if the number of degrees of freedom is large) 
and $f(\Lambda,E)$ a rapidly decreasing function of $E$, 
then to compute the number of states
with this regulating scheme one has to preform a saddle point approximation to find a saddle point
$\bar{E}$, and the regulated number of degrees of freedom can then be defined as 
\begin{equation}
\ln \Omega(\bar{E})
\end{equation}

On the other hand the black hole is dual to the CFT at some finite temperature. The partition function
of the CFT at finite temperature is 
\begin{equation}
Z(\beta)=\int e^{-\beta E} \Omega(E) dE
\label{partitionf}
\end{equation}

We can now view equation (\ref{partitionf}) in another manner. 
$Z(\beta)$ can be viewed as a regulated number of states with a particular 
regulator(cut-off). Namely
$e^{-\beta E}$ is just a cut-off function of the states of the CFT. 
Then the regulated number of degrees of freedom $\ln \Omega(\bar{E})$ 
is of course just the definition of the entropy if one views $Z(\beta)$ as a 
finite temperature
 partition function.
This suggests that the appropriate definition of the cut-off CFT is with an 
exponential decreasing
 regulating function.
The question is, is this compatible with other properties of the AdS.

\section{Gravity side}
In this section we will show that the $AdS/CFT$ in the regime of semiclassical
 supergravity
suggests a regularization scheme one should use, which matches
the one from the previous section.

\subsection{Cut-off and lightcone}
The relation between the energy cut-off and the radial coordinate of the hyper-surface
on which the cut-off theory is defined, is encoded in
the null geodesic equation. A localized  object in the bulk is 
represented by a blob on the boundary. let us denote the size of the blob
as $L$. The Lorentzian nature of the boundary theory implies (where $t$ is the 
boundary time)
\be
\frac{dL}{dt} \leq 1
\label{lightcone}
\ee
Let us now imagine the object moving radially on some trajectory
where the radial position is changing (and hence the cut-off $\Lambda$, where
$\Lambda^{-1}$ is the energy cut-off)), then
\be
\frac{dL}{dt}=\frac{dL}{d \Lambda}\frac{d\Lambda}{dt}.
\ee
Now since the blob size changes due to the changing cut-off and in order
 not to violate equation (\ref{lightcone}) one has
\be
\frac{dL}{d\Lambda}\leq 1 \ \ \ \frac{d\Lambda}{dt}\leq 1.
\label{rest1}
\ee
Since equation (\ref{lightcone}) can be saturated this implies that the 
lightcone condition on the boundary is related to the lightcone condition
in the bulk \cite{kl1}. Since $\Lambda$ is a function of the radial direction
(labeled by $r$) this gives the bulk lightcone condition in the form  
\be
|\frac{d\Lambda(r)}{dt}| =1.
\label{cond1}
\ee

Let us now look at some examples.

We start with the metric
\be
ds^2=\frac{l^2}{z^2}(-dt^2+dx_{i}^{2}+dz^2)
\ee
The lightcone condition is
\be
\frac{dz}{dt}=1
\ee
which then gives 
\be
\Lambda=z
\ee

In another form we can take
\be
ds^2=\frac{U^2}{R^2}(-dt^2+dx_{i}^{2})+\frac{R^2}{U^2}dU^2
\ee
The lightcone condition is
\be
\frac{dU}{dt}=\frac{U^2}{R^2}
\ee
To turn this equation into the form of (\ref{cond1}) one gets 
\be
\Lambda=|\int R^2 \frac{dU}{U^2}|= \frac{R^{2}}{U}.
\ee
which agrees with $\cite{pp}$. 
One can also take the metric for global AdS, of the form
\be
ds^2=-(1+\frac{r^2}{l^2})dt^2+(1+\frac{r^2}{l^2})^{-1}dr^2+r^2d\Omega^{2}
\ee
Here we find
\be
\Lambda=l\int\frac{dr}{1+\frac{r^2}{l^2}}=l(\arctan (\frac{r}{l}) +c).
\ee
We need to obey the boundary condition $\Lambda(r=\infty)=0$, so we find
\be
r=l\tan(-\frac{\Lambda}{l}+\frac{\pi}{2})
\ee
Notice that this implies that at $r=0$ the energy cut-off is
$\frac{2}{l\pi}$ which is appropriate for a theory on a sphere.

\subsection{Regulating function}

For simplicity we work with the Euclidean metric 
\begin{equation}
ds^{2}=\frac{l^2}{z^2}(dz^2 +dx_{i}^{2}).
\end{equation}
 In the semiclassical gravity regime a scalar field in the bulk is related to 
a scalar
field at the boundary as
\be
\phi(x,z)=\int dx' G(z,x,x')\phi(x')
\label{regop}
\ee
Where the kernel depends on the mass of the scalar (or on the conformal 
dimension of the corresponding CFT operator $\Delta$), and is given by 
\cite{gkpw}
\be
G(z,x,x')\sim (\frac{z}{z^2-|x-x'|^2})^{\Delta}.
\ee
Since the supergravity fields are dual to sources that couple to operators in 
the CFT,
We can think of equation (\ref{regop}) as defining a smeared source
and thus a regularization scheme\footnote{In general bulk operators are not 
simply related to boundary operators \cite{lp} other than in the semiclassical
 regime.}. To see more clearly the content of this
scheme we Fourier transform equation (\ref{regop})
\be
\phi(\vec{k})\sim K_{\Delta-d/2}(kz)\phi'(\vec{k})
\label{regfor}
\ee
where $K_{\nu}(x)$ is a  Bessel function and $k=|\vec{k}|$. The asymptotic 
behavior of
the Bessel function is (up to powers of $kz$ which are not relevant for the saddle 
point) 
\begin{equation}
\sim e^{-kz}
\end{equation} 
for all $\nu$.
Similar regularization scheme can be found for vector and higher spin fields. 
The 
important thing is that they all have the same exponential asymptotic form at
 large $kz$.
So we see that indeed the $AdS/CFT$ relationship predicts a regulating 
function of the form
that is compatible with the relationship to the thermal partition function of the CFT and thus to the black hole 
entropy\footnote{The saddle point is expected to be at $\bar{{\bf P}}=0$ 
where ${\bf P}$ is the momenta}, with a cutoff
\begin{equation}
z \sim 1/T.
\label{st}
\end{equation}

To see if this cut-off relationship is correct, all we have to do is to 
check if the radius size of 
a black hole at temperature $T$ obeys equation (\ref{st}).

Indeed for $AdS_{3}$ we have 
\be
z_{h}=\frac{1}{2\pi T}
\ee
and for $AdS_{5}$ one gets
\be
z_{h}=\frac{1}{\pi T}
\ee
Note that this does not follow from dimensional analysis since any 
function of the dimensionless coupling constant could have been involved.
Now we see that approximating the regularization function by its asymptotic behavior is valid since in the semiclassical regime $\bar{E}/T \sim S_{bh} \gg 1$.

\subsection{Non conformal case}

For non conformal theories which are dual to the near horizon geometry
of p-branes, one can do a similar analysis.
The geometry is given by \cite{imsy}
\bea
ds^2=\frac{U^{(7-p)/2}}{\sqrt{d_{p}e^2}}(-dt^2+dx_{i}^{2})+
\frac{\sqrt{d_{p}e^2}}{U^{(7-p)/2}}dU^2 +\sqrt{d_{p}e^2} U^{(p-3)/2} 
d\Omega_{8-p}^{2}
\eea
where $e^2=g_{ym}^{2}N$ and 
$d_{p}=2^{7-2p}\pi^{\frac{9-3p}{2}}\Gamma(\frac{7-p}{2})$.
The wave equation for a massless scalar in ten dimension is given by 
(see for example \cite{go})
\begin{equation}
[\partial_{U}(U^{8-p}\partial_{U})-\sqrt{d_{p}e^2} k^2 U +l(l+7-p) U^{6-p}]\phi=0
\label{gp}
\end{equation}
where $l(l+7-p)$ is some eigenvalue of the Laplacian on the sphere $S^{8-p}$.
If we define
\begin{equation}
z=\frac{\sqrt{d_{p}e^2}}{\frac{5-p}{2}U^{(5-p)/2}}
\end{equation}
Then the appropriate solution takes the form
\begin{equation}
(z)^{(p-7)/(p-5)}K_{\nu}(zk)
\end{equation}
Since Solution of equation (\ref{gp}) controls the behavior of the bulk to 
boundary Greens function at large $kz$, we see that the behavior at large 
$kz$ is again
\begin{equation}
 e^{-kz}.
\end{equation}

For this to match the gauge theory thermal entropy one should then have
that a black hole of temperature $T$ have a horizon size
\begin{equation}
T \sim 1/z_{h}
\end{equation}
which is indeed the case  since for these black holes
\be
T=\frac{7-p}{2\pi(5-p)}\frac{1}{z_{h}}
\ee

To conclude, we see that the thermal entropy relationship with the number of degrees of freedom
of a cut-off gauge theory is based on a certain regulating scheme whose functional form (in the large $N$ large t Hooft coupling region) can be predicted from the semiclassical gravity. 

\section{Acknowledgments}
I would like to thank D. Kabat for many helpful discussions.



\begin{thebibliography}{99}
\bibitem{bh}
J.~D.~Bekenstein,
{\sl Black holes and entropy,}
Phys.\ Rev.\ D {\bf 7}, 2333 (1973); \\
S.~W.~Hawking,
{\sl Particle creation by black holes,}
Commun.\ Math.\ Phys.\  {\bf 43}, 199 (1975).

\bibitem{hol}
G.~'t Hooft,
arXiv:gr-qc/9310026.\\
C.~R.~Stephens, G.~'t Hooft and B.~F.~Whiting,
Class.\ Quant.\ Grav.\  {\bf 11}, 621 (1994)
[arXiv:gr-qc/9310006].\\
L.~Susskind,
J.\ Math.\ Phys.\  {\bf 36}, 6377 (1995)
[arXiv:hep-th/9409089].

\bibitem{mald}
J.~M.~Maldacena,
Adv.\ Theor.\ Math.\ Phys.\  {\bf 2}, 231 (1998)
[Int.\ J.\ Theor.\ Phys.\  {\bf 38}, 1113 (1999)]
[arXiv:hep-th/9711200].

\bibitem{suswit}
L.~Susskind and E.~Witten,
{\sl The holographic bound in anti-de Sitter space},
arXiv:hep-th/9805114.

\bibitem{ikll2}
N.~Iizuka, D.~Kabat, G.~Lifschytz and D.~A.~Lowe,
arXiv:hep-th/0212246.

\bibitem{kl1}
D.~Kabat and G.~Lifschytz,
JHEP {\bf 9905}, 005 (1999)
[arXiv:hep-th/9902073].

\bibitem{pp}
A.~W.~Peet and J.~Polchinski,
Phys.\ Rev.\ D {\bf 59}, 065011 (1999)
[arXiv:hep-th/9809022].

\bibitem{gkpw}
S.~S.~Gubser, I.~R.~Klebanov and A.~M.~Polyakov,
Phys.\ Lett.\ B {\bf 428}, 105 (1998)
[arXiv:hep-th/9802109].\\
E.~Witten,
Adv.\ Theor.\ Math.\ Phys.\  {\bf 2}, 253 (1998)
[arXiv:hep-th/9802150].

\bibitem{lp}
G.~Lifschytz and V.~Periwal,
JHEP {\bf 0004}, 026 (2000)
[arXiv:hep-th/0003179].

\bibitem{imsy}
N.~Itzhaki, J.~M.~Maldacena, J.~Sonnenschein and S.~Yankielowicz,
{\sl Supergravity and the large N limit of theories with sixteen supercharges,}
Phys.\ Rev.\ D {\bf 58}, 046004 (1998)
[hep-th/9802042].

\bibitem{go}
T.~Gherghetta and Y.~Oz,
Phys.\ Rev.\ D {\bf 65}, 046001 (2002)
[arXiv:hep-th/0106255].

\end{thebibliography}
\end{document}